# The Cyclic Universe

Ding-Yu Chung

The cyclic universe model is a modification of the ekpyrotic universe and the pyrotechnic universe models. The cosmological model is the cyclic universe based on the cosmic cycle of the fractionalization and condensation. The cyclic universe goes through the six transitions: the triplet universe, the inflation, the big bang, the quintessence, the big crush, and the deflation transitions. The universe starts with eleven dimensional space-time with two boundary 9-branes (the pre-observable boundary 9-brane and the hidden boundary 9-brane) separated by a finite gap spanning an intervening bulk volume. The triplet transition starts when the bulk 9-brane is generated from the hidden boundary 9-brane, and collides with the pre-observable 9-brane. The collision starts the inflation transition. The collision is the brane dimensional interference mixing between the pre-observable 9-brane and the bulk 9-brane. The results are the mixed branes (combined brane dimensions) from 3- mixed brane to 9-mixed brane, the internal space (cancelled brane dimensions) where gauge fields and matter fields reside, the bulk space where gravity resides, 3-brane vacuum, and cosmic radiation. The brane dimensional interference mixing ends with equal mass proportions of mixed branes from 3 to 9 (ordinary and exotic dark matters). All mixed branes formed in the brane dimensional interference mixing occur simultaneously, so superluminal expansion (the inflation) is necessary to accommodate simultaneous brane dimensional interference mixing of all high and low energy mixed branes. Cosmic radiation generated during the inflation leads to the big bang for the cosmic expansion. Meanwhile, the hidden brane undergoes stepwise fractionalization, changing in stepwise manner from 9-brane to 3-brane. The observable universe expands in a constant rate until the quintessence transition, involving the interaction between the hidden universe and the observable universe, and causing the late cosmic accelerating expansion and contraction in the observable universe. After the quintessence transition, the observable universe starts to contract in a normal rate. Afterward, there are the big crush transition (the reverse of the big bang) and the deflation (the reverse of the inflation). The cosmic structure again consists of eleven dimensional space-time with two boundary 9-branes separated by a finite gap spanning an intervening bulk volume. The 3-mixed brane is the mixture of leptons and quarks. The structure of 3-mixed brane with brane space dimensions in internal space and bulk space resembles to the structure of atomic orbital. Consequently, the periodic table of elementary particles can be constructed to account of all ordinary leptons, quarks, gauge bosons, and hadrons. The masses of ordinary leptons, quarks, gauge bosons, and hadrons can be calculated with only four known constants: the number (seven) of spatial dimensions in the internal space and the bulk space, the mass of electron, the mass of $Z°$, and $\alpha_e$. The calculated masses are in good agreement with the observed values.

## *1. Introduction*

Both of the ekpyrotic universe model [1] by J. Khoury, B. A. Ovrut, P. J. Steinhardt, and N. Turok and its modification, the pyrotechnic universe model [2], by R. Kallosh, L. Kofman, and A. Linde involve a five-dimensional space-time with two boundary 3-branes separated by a finite gap spanning an intervening bulk volume. The two boundary 3-branes are the visible brane (our observed universe) and the hidden brane. The hidden brane occupies larger volume than the visible branes. The bulk volume contains an additional 3-brane, which is free to move across the bulk space.

The hot big bang is created by the collision of the slowly moving bulk brane with our observable brane. The energy from the collision is translated into matter and radiation, heating the universe a temperature a few orders of magnitude smaller than the unification scale. The gravitational backreaction due to the kinetic energy of the bulk brane triggers cosmic expansion. As pointed out in [2], a plausible solution for the pyrotechnic universe model is inflation.

The cyclic universe model is a modification of the ekpyrotic universe and the pyrotechnic universe models. The modification involves the replacement of two boundary 3-branes and one bulk 3-brane in a five-dimensional space-time in the ekpyrotic universe and the pyrotechnic universe model by the two boundary 9-branes (the pre-observable boundary 9-brane and the hidden boundary 9-brane) and one bulk 9-brane in eleven-dimensional space-time. As in the pyrotechnic universe model, there is inflation. The cosmological model is the cyclic universe based on the cosmic cycle of the fractionalization and condensation. The cyclic universe goes through the six transitions: the triplet universe, the inflation, the big bang, the quintessence, the big crush, and the deflation transitions (Fig. 1).

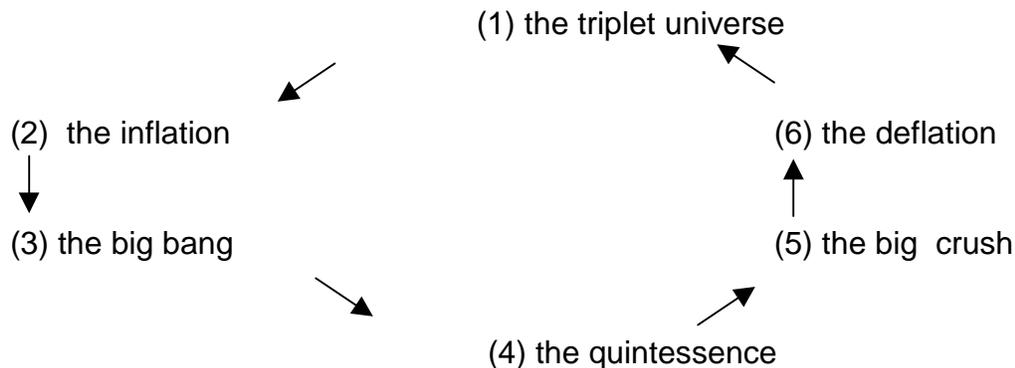

**Fig. 1**: the six cosmic transitions

The universe starts with eleven dimensional space-time with two boundary 9-branes (the pre-observable boundary 9-brane and the hidden boundary 9-brane) separated by a finite gap spanning an intervening bulk volume. The triplet transition starts when the bulk 9-brane is generated from the hidden boundary 9-brane. As soon as the bulk 9-brane is generated, it quickly collides with the pre-observable 9-brane. The collision starts the inflation transition. During the inflation, there is a continuos flow of the bulk 9-brane generated from the hidden brane to maintain a continuos collision.

The collision is the brane dimensional interference mixing between the pre-observable 9-brane and the bulk 9-brane. Some of the brane space dimensions of the bulk branes combine with the brane space dimensions of the pre-observable branes, while some of them cancel the brane space dimensions of the pre-observable branes. The combined branes are the "mixed branes", consisting of mixed branes from 3- to 9- mixed branes with 3- to 9- brane space dimensions. The space for the canceled brane space dimensions is internal space where gauge fields and matter fields reside. Each mixed brane still has one bulk space where gravity resides as in the Randall - Sundrum mechanism [3]. The brane dimensional interference mixing is shown in Fig. 2.

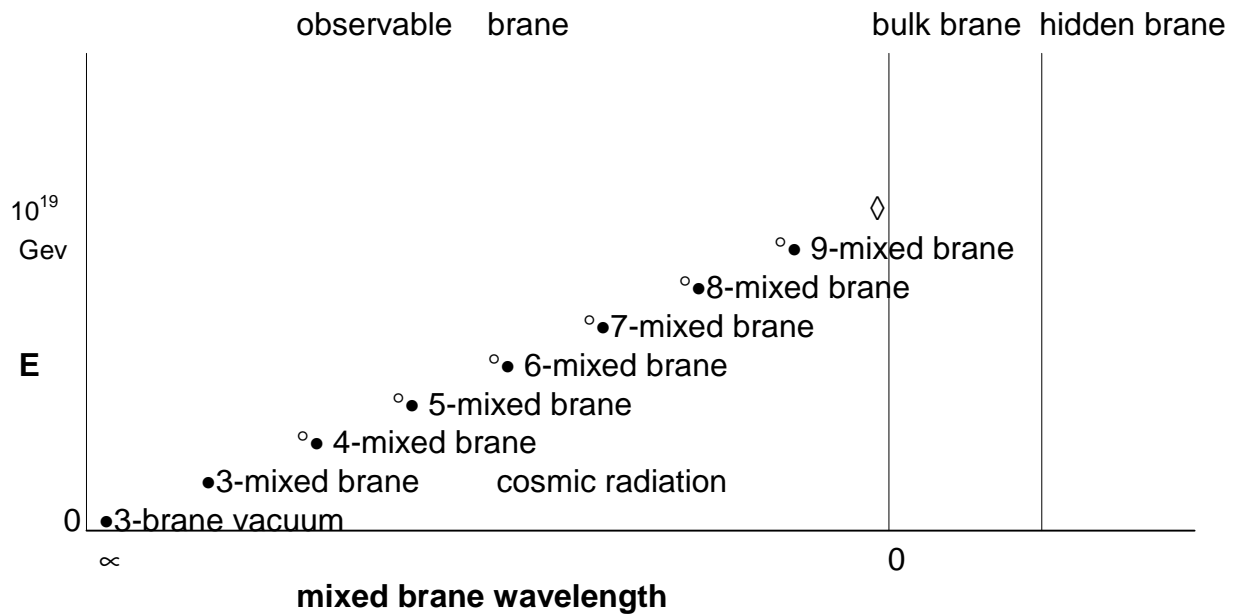

**Fig 2**: the inflation: the simultaneous brane dimensional interference mixing
• = mixed brane , ○ = internal space, ◊ = bulk space
3-mixed brane = ordinary (baryonic) matter
4- to 9- mixed branes = exotic dark matter

In this proposal, energy increases with dimensionality in a cascade manner. (A higher dimensional space-time has higher energy than the lower dimensional space-time.) The lower brane has less energy than the higher brane. Consequently, the brane dimensional interference mixing results in the release of enormous energy as cosmic radiation residing in the four space-time dimensional vacuum (3-brane vacuum). Therefore, the energy resulted from the collision between the pre-observable 9-brane and the bulk brane is in the form of the energy release from the brane dimensional interference mixing.

The ordinary (baryonic) matter is 3-mixed brane (the mixture of lepton and quarks), consisting of one bulk space, 6 internal space dimensions, and cosmic radiation in 3-brane vacuum space. The brane dimensional interference mixing transforming 9-brane to n-mixed brane is as follows.

(9 - n) brane space dimensions in the pre-observable brane + bulk brane
= n-mixed brane with n brane space dimensions + bulk space + (9 - n) internal space dimensions + cosmic radiation in 3-brane vacuum

This brane dimensional interference mixing is essentially the fractionalization of the pre-observable boundary 9-brane and the bulk 9-brane. The brane dimensional interference mixing ends with equal mass proportions of mixed branes from 3 to 9 (ordinary and exotic dark matters). All mixed branes formed in the brane dimensional interference mixing occur simultaneously, so superluminal expansion (the inflation) is necessary to accommodate simultaneous brane dimensional interference mixing of all high and low energy mixed branes. Therefore, as in the pyrotechnic universe model, this cycle universe model goes through the inflation transition.

Cosmic radiation generated during the inflation leads to the big bang for the cosmic expansion in the observable universe. After the brane dimensional interference mixing, the space-time in the hidden brane and the space-time in the observable brane are no longer the same, so the bulk space ceases to exist, and the remaining bulk brane that does not involve in the collision becomes a part of the hidden brane.

Meanwhile without the mixing of branes, the hidden brane undergoes in a different kind of expansion: stepwise fractionalization. In stepwise fractionalization, the hidden brane changes in stepwise manner from 9-brane to 3-brane, followed by condensation from 3-brane to 9-brane later. The stepwise fractionalization and condensation lead to a very slow steady expansion and contraction without cosmic radiation. The hidden brane does not have gauge force fields and matter fields.

The observable universe expands in a constant rate until the quintessence transition, when the hidden brane fractionalizes into 3-brane, compatible with the 3-brane vacuum in the observable universe. The compatibility allows the hidden brane again generates the bulk brane as quintessence that moves to the 3-brane observable vacuum in the observable universe, causing the late cosmic accelerating expansion in the observable universe. After a certain period, the cosmic cycle of franctionalization-condensation starts the condensation phase. The quintessence bulk 3-brane starts to condense into 4-brane, incompatible with the 3-brane vacuum in the observable universe. Consequently, quintessence in the observable universe moves back to the bulk space, causing the cosmic accelerating contraction in the observable universe. The quintessence transition involves both the accelerating expansion and the accelerating contraction for the observable universe. It is also the tuning point for the hidden universe and the quintessence bulk universe from the stepwise fractionalization to the stepwise condensation.

When all quintessence moves back to the bulk space, the observable universe and hidden universe do not have the compatible space-time, so the bulk space ceases to exist, and the quintessence bulk brane becomes a part of the hidden universe. The observable universe starts to contract in a normal rate. At the end of the contraction, the observable universe becomes essentially a cosmic black hole, and the hidden brane approaches to become 9-brane.  This is the big crush transition, the reverse of the big bang.

The deflation transition, the reverse of the inflation for the observable universe, occurs when there is simultaneous condensation process.  All lower branes condense to pre-observable 9-branes and bulk 9-branes that return to the hidden branes.  Meanwhile, the branes in the hidden universe reaches 9-branes through the stepwise condensation.  The cosmic structure again consists of eleven dimensional space-time with two boundary 9-branes separated by a finite gap spanning an intervening bulk volume.  A new cosmic cycle starts with the flow of the bulk 9-brane from the hidden 9-brane through the bulk space to the pre-observable brane.  The universe is a continuous cycle of fractionalization and condensation along dimensionality.

The 3-mixed brane is the mixture of leptons and quarks.  The structure of 3-mixed brane with space dimensions in internal space and bulk space resembles to the structure of atomic orbital.  Consequently, the periodic table of elementary particles can be constructed to account of all ordinary leptons, quarks, gauge bosons, and hadrons.  The masses of ordinary leptons, quarks, gauge bosons, and hadrons can be calculated with only four known constants: the number (seven) of spatial dimensions in the internal space and the bulk space, the mass of electron, the mass of Z°, and $\alpha_e$.  The calculated masses are in good agreement with the observed values.

## 2.   The Dimension Hierarchy

The universe is a continuous cycle of fractionalization and condensation along dimensionality.  In this proposal, energy increases with dimensionality in a cascade manner.  Each space dimension can be described by a fermion and a boson as the following hierarchy [5]:

$B_4$ $F_5$ $B_5$ $F_6$ $B_6$ $F_7$ $B_7$ $F_8$ $B_8$ $F_9$ $B_9$ $F_{10}$ $B_{10}$ $F_{11}$ $B_{11}$

where B and F are boson and fermion in each space-time dimension. The probability to transforming a fermion into its boson partner in the adjacent dimension is same as the fine structure constant, $\alpha$, the probability of a fermion emitting or absorbing a boson.  The probability to transforming a boson into its fermion partner in the same dimension is also the fine structure constant, $\alpha$.  This hierarchy can be expressed in term of the dimension space-time number, D,

$$M_{D-1, B} = M_{D, F} \alpha_{D, F}, \quad (1)$$

$$M_{D, F} = M_{D, B} \alpha_{D, B}, \quad (2)$$

where $M_{D, B}$ and $M_{D,F}$ are the masses for a boson and a fermion, respectively, and $\alpha_{D, B}$ or $\alpha_{D,F}$ is the fine structure constant, which is the ratio between the energies of a boson and its fermionic partner.  All fermions and bosons are related by the order $1/\alpha$. Assuming $\alpha_{D,B} = \alpha_{D,F}$, the relation between the bosons in the adjacent dimensions, then, can be expressed in term of the space-time dimension number, D,

$$M_{D-1, B} = M_{D, B} \, \alpha^2_D , \qquad (3)$$

## 3. *The Matter Fields and the Gauge Fields in the Ordinary Universe*

The collision of the bulk 9-brane and the pre-observable boundary 9-brane leads to the observable universe consisting of the mixed branes.  Our observable ordinary (baryonic) universe is 3-mixed brane.  Exotic dark matter in the observable universe has various branes from 4-mixed brane to 9-mixed brane.  It does not have the same gauge fields as the ordinary matters, but it has the same gravity, so it cannot be seen, but can be observed by gravity. The ordinary (baryonic) matter is one of the seven mixed branes at equal mass proportions, so the baryonic mass fraction is calculated to be 0.14.  The universal baryonic mass fraction was found to be 0.13 by the observations of primordial deuterium abundance [4].   The calculated value agrees well with the observed value.

Before the brane dimensional interference mixing, the bulk brane has 10 space dimensions, and the pre-observable brane has 10 space dimensions, so the total number of space dimension is 20.  In order to maintain the conservation of number of space dimensions (brane space, internal space, and bulk space dimensions), the structure of the observable branes after the brane dimensional interference mixing has two separate sets of 10 space dimensions.  One set is the main set with 10 space dimensions, and the other set is the auxiliary set with 10 space dimensions.  The auxiliary set is dependent on the main set, and the auxiliary set is derived from the main set, so the total set appears to have 10 space dimensions.  For ordinary matter, the main set consists of the lepton 3-brane and the 7 main space dimensions in internal-bulk space.  The auxiliary set consists of the quark 3-brane, and the 7 auxiliary space dimensions in internal-bulk space.  There are two separate sets of 10 space dimensions, and the total number of space dimensions is 20.  The total set is shown in Fig. 3.

Since the lepton 3-brane is the main brane, and the quark 3-brane is the auxiliary brane, quarks must behave as leptons (integer charge and hypercharge) in order to be observable.  The seven main dimensions are for all major functions, including light leptons, dimensional fermions, gauge bosons, and gravity.   The seven auxiliary dimensions are for mostly quarks, including quark matter field, heavy lepton (μ and τ) matter field, and the binding energy among quarks in hadrons.

As shown in Fig. 3, in the internal-bulk space, the auxiliary space dimensions locate in the middle of the main space dimensions.  In internal-bulk space, the seventh main space-time dimension is marked as the start of the auxiliary space dimensions as the fifth space-time dimension is the start of the main space dimensions, because there is a mixing (as in the symmetry mixing in the electroweak interaction between U(1) and SU(2)) between the fifth and the seventh main space-time dimensions. The first auxiliary dimension is parallel to the seventh main space-time dimension.

The structure of 3-mixed brane with space dimensions in internal space and bulk space resembles to the structure of atomic orbital. Consequently, the periodic table of elementary particles can be constructed to account of all ordinary leptons, quarks, gauge bosons, and hadrons as described in details in Reference 5. It is briefly reviewed here.

Each of six main spatial dimensions in internal space is represented by a gauge field with a specific internal symmetry for specific space-time symmetry. CP nonconservation is required to distinguish the permanent lepton-quark composite state from the CP symmetrical radiation that is absence of the permanent lepton-quark composite state. P nonconservation is required to achieve chiral symmetry for massless leptons (neutrinos), so some of the dimensional fermions can become leptons (neutrinos).

Various internal symmetry groups in internal-bulk space are require to organize leptons, quarks, black hole particles, and force fields. The force fields include the long-range massless force to bind leptons and quarks, the short-range force to bind quarks, and the short-range interactions for the flavor change among quarks and leptons.

Finally, the ordinary gauge bosons, leptons, and quarks absorb the common scalar field of four-dimensional space-time to acquire the common label of four-dimensional space-time through Higgs mechanism. The results are the ordinary gauge bosons (Table 1) and the periodic table of ordinary elementary particles (Table 2).

For the gauge bosons, the seven main dimensions in internal-bulk space are arranged as the below.

$$F_5 \ B_5 \ F_6 \ B_6 \ F_7 \ B_7 \ F_8 \ B_8 \ F_9 \ B_9 \ F_{10} \ B_{10} \ F_{11} \ B_{11}$$

where B and F are boson and fermion in each space-time dimension. The ordinary gauge bosons can be derived from Eqs. (1) and (2). Assuming $\alpha_{D,B} = \alpha_{D,F}$, the relation between the bosons in the adjacent dimensions, then, can be expressed in term of the dimension number, D,

$$M_{D-1, B} = M_{D, B} \ \alpha^2_D , \qquad (3)$$

where D= 6 to 11, and $E_{5,B}$ and $E_{11,B}$ are the energies for the dimension five and the dimension eleven, respectively.

The lowest energy is the Coulombic field, $E_{5,B}$

$$\begin{aligned} E_{5, B} &= \alpha \ M_{6,F} \\ &= \alpha \ M_e , \end{aligned} \qquad (4)$$

where $M_e$ is the rest energy of electron, and $\alpha = \alpha_e$, the fine structure constant for the magnetic field. The bosons generated are called "dimensional bosons" or "$B_D$". Using only $\alpha_e$, the mass of electron, the mass of $Z^0$, and the number (seven) of spatial dimensions in internal-bulk space, the masses of $B_D$ as the ordinary gauge boson can be calculated as shown in Table 1.

**Table 1.** The Energies of the Dimensional Bosons
$B_D$ = dimensional boson, $\alpha = \alpha_e$

| $B_D$ | $M_D$ | GeV | Ordinary gauge boson | Interaction, symmetry |
|---|---|---|---|---|
| $B_5$ | $M_e \alpha$ | $3.7 \times 10^{-6}$ | A | electromagnetic, U(1) |
| $B_6$ | $M_e/\alpha$ | $7 \times 10^{-2}$ | $\pi_{1/2}$ | strong, SU(3) |
| $B_7$ | $M_6/\alpha_w^2 \cos\theta_w$ | 91.177 | $Z_L^0$ | weak (left), SU(2)$_L$ |
| $B_8$ | $M_7/\alpha^2$ | $1.7 \times 10^6$ | $X_R$ | CP (right) nonconservation, U(1)$_R$ |
| $B_9$ | $M_8/\alpha^2$ | $3.2 \times 10^{10}$ | $X_L$ | CP (left) nonconservation, U(1)$_L$ |
| $B_{10}$ | $M_9/\alpha^2$ | $6.0 \times 10^{14}$ | $Z_R^0$ | weak (right), SU(2)$_R$ |
| $B_{11}$ | $M_{10}/\alpha^2$ | $1.1 \times 10^{19}$ | G | gravity, space-time |

In Table 1, $\alpha_w$ is not same as $\alpha$ of the rest, because there is a mixing between $B_5$ and $B_7$ as the symmetry mixing between U(1) and SU(2) in the standard theory of the electroweak interaction, and $\sin\theta_w$ is not equal to 1. As shown in Reference 5, $B_5$, $B_6$, $B_7$, $B_8$, $B_9$, and $B_{10}$ are A (massless photon), $\pi_{1/2}$, $Z_L^0$, $X_R$, $X_L$, and $Z_R^0$, respectively, responsible for the electromagnetic field, the strong interaction, the weak (left handed) interaction, the CP (right handed) nonconservation, the CP (left handed) nonconservation, and the P (right handed) nonconservation, respectively. The calculated value for $\theta_w$ is $29.69^0$ in good agreement with $28.7^0$ for the observed value of $\theta_w$ [6]. The calculated energy for $B_{11}$ is $1.1 \times 10^{19}$ GeV in good agreement with the Planck mass, $1.2 \times 10^{19}$ GeV.

The calculated masses of all ordinary gauge bosons are in good agreement with the observed values. Most importantly, the calculation shows that exactly seven main spatial dimensions in internal and bulk space are needed for all fundamental interactions.

The two sets of seven spatial dimensions in internal-bulk space result in 14 spatial dimensions (Fig. 3) for gauge bosons, leptons, and quarks. Quarks and heavy leptons ($\mu$ and $\tau$) are in seven auxiliary spatial dimensions. The periodic table for ordinary elementary particles is shown in Table 2.

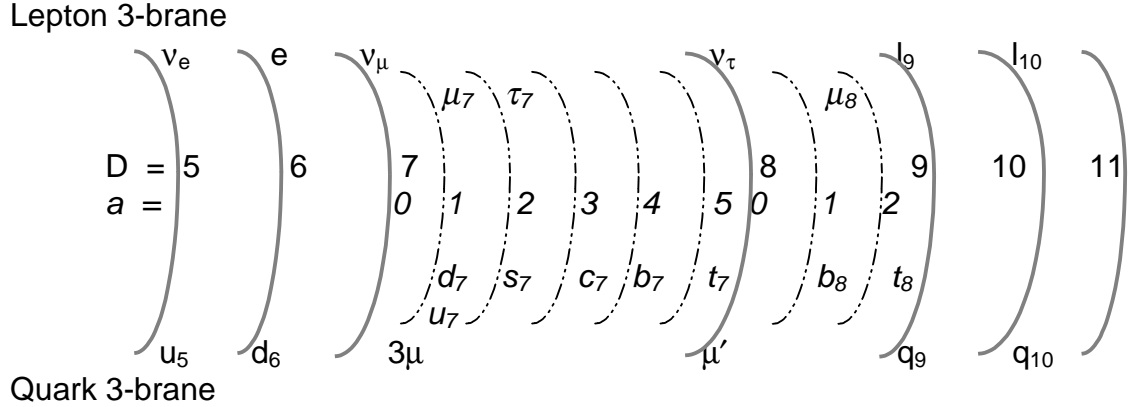

Lepton 3-brane

Quark 3-brane

**Fig. 3.** Ordinary leptons and quarks in the dimensional orbits
D = main dimensional number, a = auxiliary dimensional number

**Table 2.** The Periodic Table of ordinary elementary particles
D = dimensional number, a = auxiliary dimensional number

| D | a = 0 | 1 | 2 | a = 0 | 1 | 2 | 3 | 4 | 5 | |
|---|---|---|---|---|---|---|---|---|---|---|
|  | Lepton |  |  | Quark |  |  |  |  |  | Boson |
| 5 | $l_5 = \nu_e$ |  |  | $q_5 = u_5 = 3\nu_e$ |  |  |  |  |  | $B_5 = A$ |
| 6 | $l_6 = e$ |  |  | $q_6 = d_6 = 3e$ |  |  |  |  |  | $B_6 = \pi_{1/2}$ |
| 7 | $l_7 = \nu_\mu$ | $\mu_7$ | $\tau_7$ | $q_7 = 3\mu$ | $u_7/d_7$ $s_7$ | $c_7$ | $b_7$ | $t_7$ |  | $B_7 = Z_L^0$ |
| 8 | $l_8 = \nu_\tau$ | $\mu_8$ |  | $q_8 = \mu'$ | $b_8$ | $t_8$ |  |  |  | $B_8 = X_R$ |
| 9 | $l_9$ |  |  | $q_9$ |  |  |  |  |  | $B_9 = X_L$ |
| 10 | $F_{10}$ |  |  |  |  |  |  |  |  | $B_{10} = Z_R^0$ |
| 11 | $F_{11}$ |  |  |  |  |  |  |  |  | $B_{11} = G$ |

     D is the dimensional orbital number for the seven main spatial dimensions in internal-bulk space. The auxiliary dimensional orbital number, a, is for the seven auxiliary spatial dimensions, mostly for subquarks. All ordinary gauge bosons, leptons, and subquarks are located on the seven dimensional orbits and seven auxiliary orbits. Most leptons are dimensional fermions, while all quarks are the sums of subquarks.
     The fermion mass formula for massive leptons and quarks is derived from Reference 5 as follows.

$$M_{F_{D,a}} = \sum M_{F_{D,0}} + M_{AF_{D,a}}$$
$$= \sum M_{F_{D,0}} + \frac{3}{2} M_{B_{D-1,0}} \sum_{a=0}^{a} a^4 \quad (5)$$
$$= \sum M_{F_{D,0}} + \frac{3}{2} M_{F_{D,0}} \alpha_D \sum_{a=0}^{a} a^4$$

Each fermion can be defined by dimensional numbers (D's) and auxiliary dimensional numbers (a's). The compositions and calculated masses of ordinary leptons and quarks are listed in Table 3.

**Table 3.** The Compositions and the Constituent Masses of Ordinary Leptons and Quarks
D = dimensional number and a = auxiliary dimensional number

|          | $D_a$                       | Composition                 | Calc. Mass       |
|----------|-----------------------------|-----------------------------|------------------|
| Leptons  | $D_a$ for leptons           |                             |                  |
| $\nu_e$  | $5_0$                       | $\nu_e$                     | 0                |
| e        | $6_0$                       | e                           | 0.51 MeV (given) |
| $\nu_\mu$| $7_0$                       | $\nu_\mu$                   | 0                |
| $\nu_\tau$| $8_0$                      | $\nu_\tau$                  | 0                |
| $\mu$    | $6_0 + 7_0 + 7_1$           | $e + \nu_\mu + \mu_7$       | 105.6 MeV        |
| $\tau$   | $6_0 + 7_0 + 7_2$           | $e + \nu_\mu + \tau_7$      | 1786 MeV         |
| Quarks   | $D_a$ for quarks            |                             |                  |
| u        | $5_0 + 7_0 + 7_1$           | $u_5 + q_7 + u_7$           | 330.8 MeV        |
| d        | $6_0 + 7_0 + 7_1$           | $d_6 + q_7 + d_7$           | 332.3 MeV        |
| s        | $6_0 + 7_0 + 7_2$           | $d_6 + q_7 + s_7$           | 558 MeV          |
| c        | $5_0 + 7_0 + 7_3$           | $u_5 + q_7 + c_7$           | 1701 MeV         |
| b        | $6_0 + 7_0 + 7_4$           | $d_6 + q_7 + b_7$           | 5318 MeV         |
| t        | $5_0 + 7_0 + 7_5 + 8_0 + 8_2$ | $u_5 + q_7 + t_7 + q_8 + t_8$ | 176.5 GeV     |

The calculated masses are in good agreement with the observed constituent masses of ordinary leptons and quarks [7,8]. The mass of the top quark found by Collider Detector Facility is 176 ± 13 GeV [7] in a good agreement with the calculated value, 176.5 GeV. The calculated masses are in good agreement with the observed values. As shown in Reference 5, the masses of hadrons can also be calculated based on binding energy derived from the auxiliary space dimensions. The calculated values for the masses of hadrons are in good agreement with the observed values. Most importantly, the calculation shows that exactly seven auxiliary spatial dimensions are needed for all leptons and quarks.

### *4.     The Quintessence*

The observable universe expands in a constant rate until the quintessence transition, when the hidden brane fractionalizes into 3-brane. During the quintessence transition, the quintessence bulk brane (quintessence) moves to the 3-brane observable vacuum in the observable universe, causing the late cosmic accelerating expansion in the observable universe.

The evidence for late-time cosmic accelerating expansion is from the recent observations of large-scale structure that suggests that the universe is undergoing cosmic accelerating expansion, and it is assumed that the universe is dominated by a dark energy with negative pressure recently [9]. The dark energy can be provided by a non-vanishing cosmological constant or quintessence [10], a scalar field with negative pressure. However, a cosmological constant requires extremely fine-tuned [11]. Quintessence requires an explanation for the late- time cosmic accelerating expansion [12]. Why does quintessence dominate the universe only recently? One of the explanations is the *k*-essence model where the pressure of quintessence switched to a negative value at the onset of matter-domination in the universe [12].

According to the proposal, this late-time cosmic accelerating expansion is caused by the quintessence, the quintessence bulk 3-brane, moving to the four dimensional observable vacuum in the observable universe. From Eqs (1) and (2) and Table 1, the energies of the five dimensional fermion and the four dimensional boson are calculated to be 2.72 x$10^{-8}$ GeV and 1.99 x$10^{-10}$ GeV, respectively. The energy of the eleven dimensional supermembrane is 1.1x$10^{19}$ GeV from Table 1. The energy ratio between the eleven dimensional supermembrane and the five dimensional fermion and the four dimensional boson are 4.17 x$10^{26}$ and 5.71 x$10^{28}$, respectively. In other words, there are 4.17 x$10^{26}$ five dimensional fermions and 5.71 x$10^{28}$ four-dimensional bosons per eleven dimensional supermembrane. The rate of the stepwise fractionalization is assumed to be 5.39 x$10^{-44}$ second, that is $T_p^{1/4}$ where $T_p$ is the Planck time. (The vacuum energy is equal to $M_p^4$ where $M_p$ is the Planck mass.) The total time for the fractionalization is 0.2 billion years to generate all five dimensional fermions, and 27.6 billion years to generate all four-dimensional bosons. Therefore, the quintessence input starts in 0.2 billion years after the Big Bang, and ends in 27.6 billions years, and quintessence input causes accelerating expansion in the observable universe.

At the end of cosmic accelerating expansion in the observable universe, the quintessence in the observable universe moves back to the quintessence bulk space, causing the cosmic accelerating contraction in the observable universe. It is the reverse quintessence transition. It is the start of the contraction for the observable universe, and the gradual condensation (stepwise changing from 3-brane to 9-brane) for the quintessence bulk brane and the hidden brane.

## 5.    *The Fractionalization-condensation in quantum mechanics*

The study of semi-conductor crystals at low temperature shows the fractionalization of electron in the form of a separation of spin and charge or literal fractionalization of electric charge [13]. H. J. Maris also found a way to break electron into quasi-electrons in liquid helium [14]. This spontaneous tendency for an object to fractionalize is the base to interpret quantum mechanics.

The three most important non-classical features in quantum mechanics are the nonlocal operation, the interference effect, and the collapse of wavefunction. In this proposal, the base of quantum mechanics is the spontaneous tendency for an object to fractionalize into quasistates and condense quasistates. This quasistate is equivalent to the eigenfunction in the wavefunction. An object with the fractionalization-condensation is equivalent to the unitary wavefunction. All quasistates during fractionalization can communicate with one another to preserve the integrity of the object regardless of distance, implying nonlocal operation.  The integrity of an object is equivalent to the unitary in the wavefunction.  In the wavefunction, all eigenfunctions can communicate with one another.  There is no time in the wavefunction at a fundamental level.   The combination of the fractionalization and the integrity leads to nonlocality in quantum mechanics.

At the same time, all quasitates during condensation overlap with one another because they are originally from one source.   The combination of the condensation and the integrity results in interference effect in quantum mechanics.

The disruption of the fractionalization-condensation is equivalent to the collapse of wavefunction.  When the fractionalization-condensation is disrupted externally by the interaction with environment  [15] or measurement, or is disrupted by internal transformation, such as decay, an object loses the fractionalization-condensation (nonlocality-interference).  Therefore, when an object is isolated and stable, the object has nonlocality-interference.   When an object is not isolated and stable, the object loses nonlocality-interference, and becomes a classical object instead of quantum object.

The cosmic origin of quantum mechanics is derived from the simultaneous fractionalization-condensation in the observable universe.  During the inflation, the pre-observable 9-branes fractionalize into various lower branes simultaneously and instantly.  During the deflation, various lower branes condense into 9-branes simultaneously and instantly.   The cosmic inflation and the cosmic deflation lead to the microscopic instant fractionalization-condensation that allows all quasistates from an object to appear simultaneously.

## *6.*    *Conclusion*

The cyclic universe is based on the ekpyrotic universe and the pyrotechnic universe models. The modification involves the replacement of two boundary 3-branes and one bulk 3-brane in a five-dimensional space-time in the ekpyrotic universe model by the two boundary 9-branes (the pre-observable boundary 9-brane and the hidden boundary 9-brane) and one bulk 9-brane in eleven-dimensional space-time.  As in the pyrotechnic universe model, there is the inflation.  Furthermore, the cosmological model is the cyclic universe based on the cosmic cycle of the fractionalization and condensation.  The cyclic universe goes through the six transitions: the triplet universe, the inflation, the big bang, the quintessence, the big crush, and the deflation.

The structure of matter derived from the cyclic universe model provides the base for the periodic table of ordinary elementary particles.  The baryonic mass fraction and the masses of ordinary elementary particles and hadrons can be calculated with only four known constants: the number (seven) of spatial dimensions in the internal space and the bulk space, the mass of electron, the mass of Z°, and $\alpha_e$.  The calculated masses are in good agreement with the observed values.

In this proposal, there is no supersymmetry breaking.  Supersymmetry itself is a cascade supersymmetry where energy increases with dimensionality in a cascade manner.  There is no compacification.  The spatial dimensions that greater than 3 exist in internal-bulk space. Instead of beginning and end, there is a continuous cosmic cycle of fractionalization and condensation.